# ENVIRONMENTAL SCIENCE: PHYSICAL PRINCIPLES AND APPLICATIONS

**A. Speranza**, Department of Mathematics and Computer Science, University of Camerino, Italy.

**V. Lucarini**, Department of Mathematics and Computer Science, University of Camerino, Italy.

**PACS:**

| | |
|---|---|
| **02.50.-r** | **Probability theory, stochastic processes, and statistics** |
| **02.60.-x** | **Numerical approximation and analysis** |
| **02.70.-c** | **Computational techniques** |
| **05.45.-a** | **Nonlinear dynamics and nonlinear dynamical systems** |
| **47.27.-I** | **Turbulent flows, convection, and heat transfer** |
| **47.32.-y** | **Rotational flow and vorticity** |
| **92.10.-c** | **Physics of the oceans** |
| **92.40.-t** | **Hydrology and glaciology** |
| **92.60.-e** | **Meteorology** |
| **92.70.-j** | **Global change** |

**Links to other articles:** *Computer simulation techniques, Numerical approximation and analysis, Nonlinear dynamics and nonlinear dynamical systems*

# Introduction

Environmental science almost invariably proposes problems of extreme complexity, typically characterized by strongly nonlinear evolution dynamics. The systems under investigation have many degrees of freedom - which makes them *complicated* – and feature nonlinear interactions of several different components taking place on a vast range of time-space scales – which makes them *complex*. Such systems evolve under the action of macroscopic driving (typically the solar heating) and modulating (*e.g.* the Earth's rotation and gravitation) agents. The most comprehensive example is the entire climatic system. In its most rigorous definition, the climatic system is constituted by four intimately interconnected sub-systems, atmosphere, hydrosphere, cryosphere, and biosphere (figure 1), and is powered by the electromagnetic radiation emitted by the Sun (figure 2). These subsystems are by themselves *complicated* and *complex* and interact nonlinearly with each other on various time-space scales.

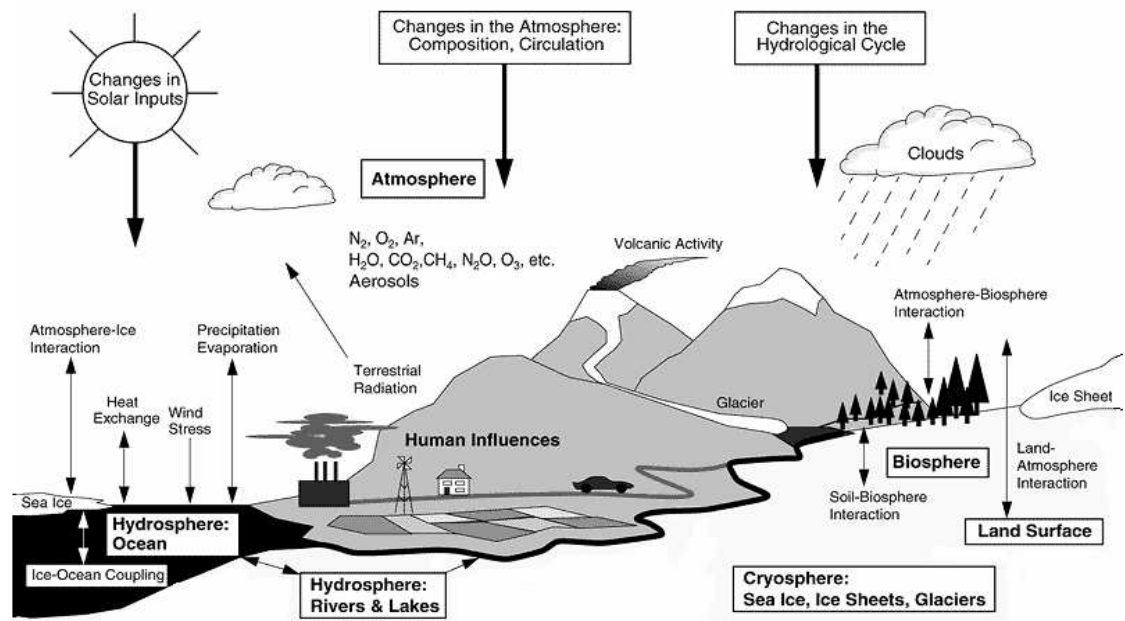

**Figure 1: Components of the climatic system; the main interactions are also indicated. From IPCC 2001.**

Moreover, environmental sciences are not usually provided with real laboratories where theories can be tested against experiments, since phenomena often take place only once and cannot be reproduced.

Theories can often be tested only against observational data from the past, which may feature problems of various degrees of criticality, essentially because of the physical extension of the systems under analysis. In many cases, the available observations feature a relatively low degree of reciprocal synchronic coherence and individually present problems of diachronic coherence, due to changes in the strategies of data gathering with time. The presence of strong variability of both the outputs of theoretical models and of the real systems contributes to blur the line between a failed and a passed test, in terms of model reliability.

It is important to emphasize that the theoretical models unavoidably suffer of two well-distinct kinds of uncertainties. The uncertainties in the initial conditions of the systems, termed *uncertainties of the first kind*, are problematic because of the effects of chaos. These uncertainties may be partially dealt with using well-suited Monte Carlo approaches on the initial conditions.

The *uncertainties of the second kind* are structural uncertainties, due to the simplifications adopted in the models for the description of the processes and of the feedbacks characterizing the system under investigation. These latter uncertainties seem to be somehow more fundamental – the dynamic framework is uncertain - and harder to deal with.

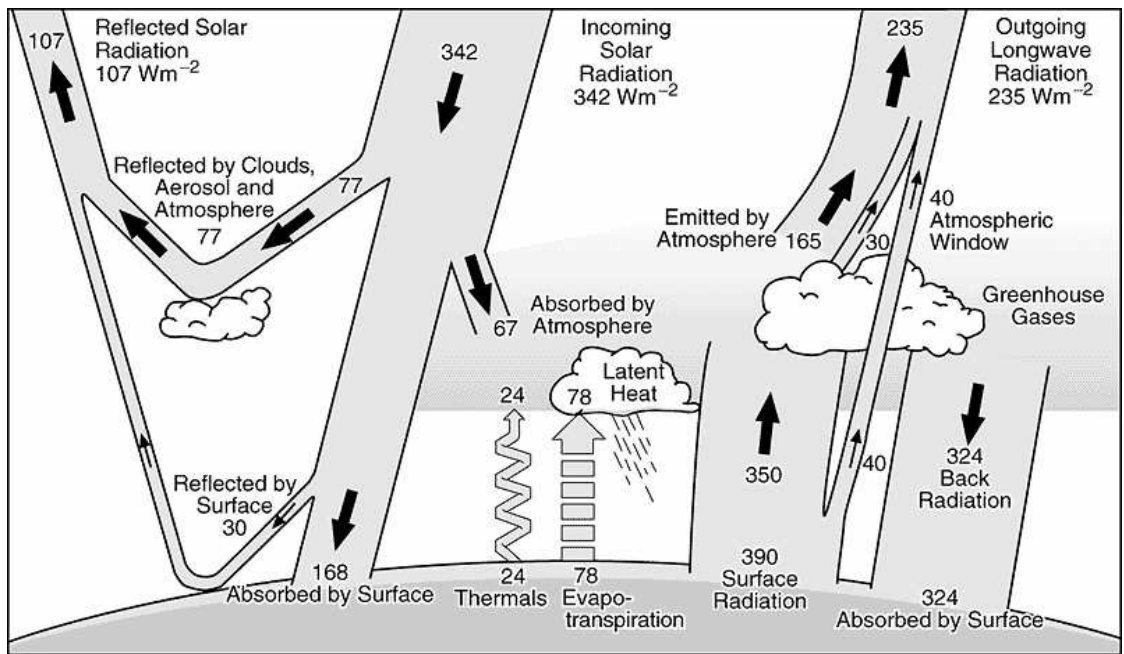

**Figure 2: Energy fluxes through the atmosphere - in units of Wm⁻². From IPCC 2001.**

Therefore, it is conceptually incorrect to expect that the essentially phenomenological theories which have been developed for the environmental sciences could provide answers having comparable precision and similar structure to those provided by theories relevant for the study of systems having lower degree of complexity.

In view of the above difficulties, it is not surprising that the historical development of Environmental Science took place essentially along two different main paths: on the one side the descriptive, basically qualitative, approach of *naturalistic* tradition, on the other the quantitative formulation of *physical-mathematical* tradition. Linneo's classification of living beings and Laplace tidal equations are classical examples. The synthesis between the two approaches took place in the nineteenth and twentieth centuries mostly through applications of fluid dynamics and/or thermodynamics to specific systems. However, a certain separation (sometimes even contradiction) between the two approaches occasionally still emerges today. Meteorological *maps*, interpreted on the basis of the analogies of the so-called *synoptic* Meteorology, are examples of the modern evolution in the descriptive tradition, while numerical weather forecasting exemplifies the quantitative one.

The description of the macroscopic dynamics of environmental systems is based on the systematic use of dominant *balances* derived on a *phenomenological* basis in order to *specialize the* dynamical equations. Such balances are suitable classes of approximate solutions of the evolution equations which represent a reasonably good approximation to the actual observed fields when sufficiently large spatial or temporal averages are considered. Actually, different balances have to be considered depending on the time and space scales we are focusing our interest on. Such an approach reflects the fundamentally heuristic-inductive nature of the scientific research in environmental sciences, where the traditional reductionistic scientific attitude is not always effective. In order to exemplify this procedure, we consider the very relevant case of the motion of the fluids that permit the existence of life on the Earth, air and water: the so-called geophysical fluids.

## Geophysical fluids and phenomenological balances

Geophysical fluid systems are very complex in microphysical structure and composition and evolve under the action of macroscopic driving (solar heating) and modulating (Earth's rotation and gravitation) agents. The complexity essentially depends on the presence of nonlinear interactions and feedbacks between the various parts of the system, which couple very different time and space scales. In many cases, the dynamics of such systems is chaotic (the self-correlation of the fields vanishes within a finite time domain) and is characterized by a large natural variability on different time scales.

The dynamics of a geophysical fluid for an observer in a uniformly rotating frame of reference is described by the Navier-Stokes equation:

(1)
$$\rho\frac{d\vec{u}}{dt} + 2\rho\vec{\Omega}\times\vec{u} = -\vec{\nabla}p - \rho\nabla\Phi + F$$

where $\rho$ is the density of the fluid, $\vec{u} = (u,v,w)$ is the velocity vector, $\vec{\Omega}$ is the Earth's rotation vector, $p$ is the pressure, $\Phi$ is the geopotential (essentially coincident with the gravitational potential), $F$ is the frictional force per unit mass, and the total derivative is expressed as:

(2)
$$\frac{d}{dt} \equiv \frac{\partial}{\partial t} + \vec{u} \cdot \vec{\nabla} \,,$$

thus including the nonlinear advection term.

Through the use of suitable approximations, the equations of motion can be focused upon the desired components of the flow. The *filtering* process is the introduction of a set of mathematical approximations into the Navier-Stokes equation having the scope of *filtering out* (excluding) solutions corresponding to the physical processes that are heuristically assumed to contribute only negligibly to the dynamics of the system at the time and space scale under examination. The magnitudes of various terms in the governing equations for a particular type of motion are estimated using the so-called scale analysis technique.

**Hydrostatic balance**

As a relevant example of these procedures, we consider the *hydrostatic approximation*. In a local Cartesian coordinate system where *z* is perpendicular to the surface an obvious fixed point (stationary solution) of the Navier-Stokes equation can be derived from the time-independent classical *hydrostatic equilibrium* equations:

$$\vec{u} = (0,0,0)$$
$$-\rho_h g - \frac{\partial p_h}{\partial z} = 0$$

(3)

where the functions $\rho_h(x,y,z,t)$ and $p_h(x,y,z,t)$ satisfying the above balance (the subscript h stands for "hydrostatic") depend only on the spatial variable *z*. Three points related to the above described balance are worth mentioning explicitly:

- the fluid velocity is assumed negligible everywhere in the system;

- the condition of hydrostatic equilibrium does not determine univocally the thermodynamic fields $\rho_h(z)$ and $p_h(z)$ but only their relationship; other information is requested in order to determine univocally the profiles of density, pressure, temperature;

- the balance is observable (and practically useful!) only if it is stable.

Now, the interesting point is that quite different geophysical flows (atmosphere, ocean, surface and, ground water, etc.) can be represented as being in near hydrostatic equilibrium when suitable spatial and temporal averaging process is considered.

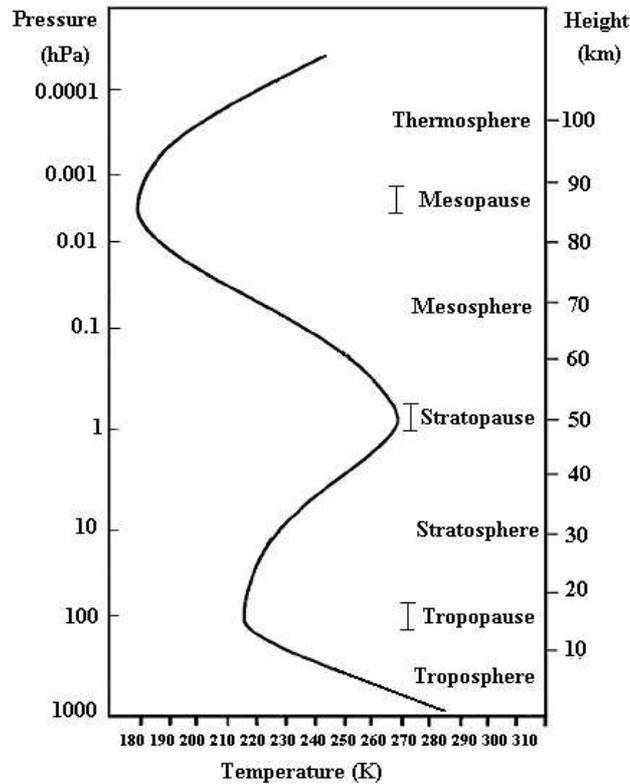

**Figure 3: Typical temperature vertical profile of the atmosphere. The main sections of the atmosphere are indicated and the small extent of the quasi-constant temperature sections is emphasized.**

In general terms, hydrostatic equilibrium is realized when the portions of a fluid with lower specific entropy are positioned below those with higher specific entropy, where directions are defined by the local gravity force vector. When this condition is broken because of an external forcing, the stratification is not stable and the fluid rearranges so that hydrostatic condition is re-established. Therefore, violations of hydrostatic equilibrium exist only on short time and space scales and often are not easily observable. Typical models are formulated in such a way that local non-hydrostatic conditions are quickly eliminated and the equilibrium condition that is recovered is *parameterized* in terms of variables explicitly represented on the numerical grids. In a later section we will briefly describe the general scope of the parameterization procedures.

As a matter of fact, non-hydrostatic models are nowadays available and are currently used. However, for any practical purpose, sufficiently detailed initial and boundary conditions are not available and the above expressed methodological problems reappear in connection with their definition

**Geostrophic balance**

Another instructive example is that of time-independent purely horizontal balanced flows where the horizontal components of the pressure gradient force and the Coriolis force cancel out in the Navier-Stokes equation. Such flows are termed *geostrophic*, which, etymologically, means Earth-turning. The structure parameter determining qualitatively the goodness of such approximation is the Rossby number $Ro=U/f_0L$ where $U$ is a characteristic horizontal velocity of the fluid, $L$ is a characteristic horizontal extension of the fluid and $f_0$ is twice the value of the projection of Earth angular velocity vector on the plane tangent to the surface of the Earth at the considered latitude. This approximation holds for relatively large scale extratropical flows in the in the regions where friction is not important – *i.e.* away from the boundaries of the fluid. Geostrophic equations are obtained as a zeroth-order $Ro$ expansion of the Navier-Stokes equation. The fluid motion is introduced by considering small perturbations breaking the translation symmetry of the baseline purely hydrostatic density and pressure fields. This is achieved by assuming that the actual density and pressure fields are given by $\rho(x,y,z)=\rho_g(x,y,z)+\rho_h(z)$ and

$p(x, y, z) = p_g(x, y, z) + p_h(z)$ respectively. Geostrophic equations thus result to be the following:

(4)
$$\rho_h u f_0 = -\frac{\partial p_g}{\partial y}$$
$$\rho_h v f_0 = \frac{\partial p_g}{\partial y}$$
$$g \rho_g = -\frac{\partial p_g}{\partial z}$$

Geostrophic and hydrostatic balance constrain atmospheric motions so that the fluid flows are uniquely determined by the pressure field, since the currents are parallel (rather then perpendicular as in non-rotating fluids) to the isobars at a given geopotential height. The time-independent nature of these equations implies that the non-geostrophic terms, although relatively small, are important for the time-evolution of the flow.

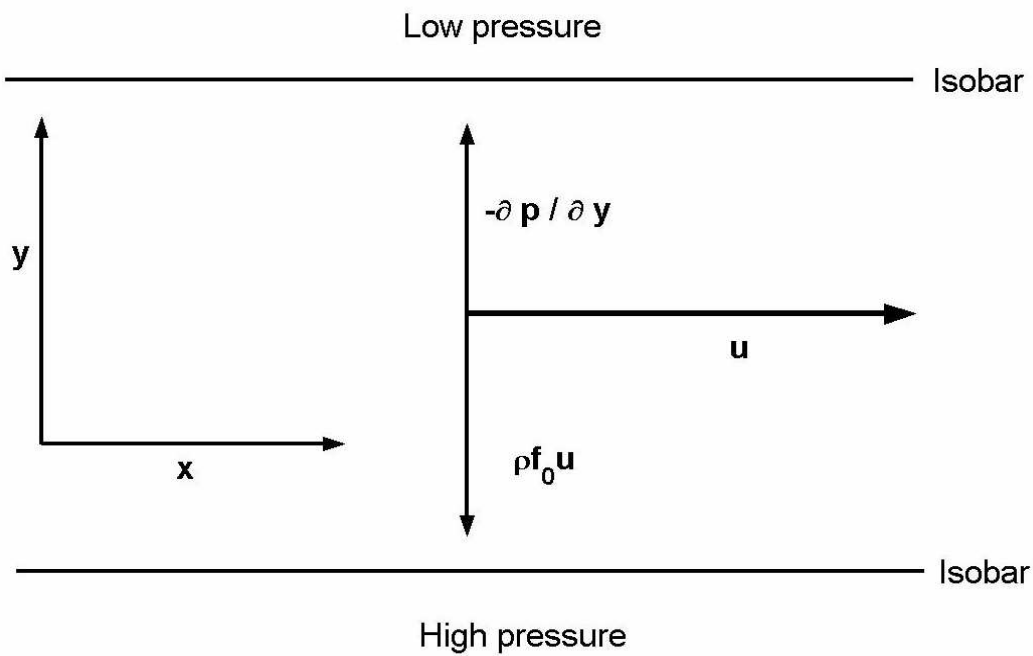

Figure 4: Geostrophic balanced flow; pressure gradient force (upper arrow) and Coriolis force (lower arrow) cancel out and the flow (horizontal arrow) is parallel to the isobars.

A system or flow that evolves slowly in time compared to $1/f_0$ can be described using the so called *quasi-geostrophic* theory. This is a perturbative theory obtained as a first-order *Ro* filtering of the Navier-Stokes equation and analyzes departures from geostrophic balance. The quasi-geostrophic approximation is used in the analysis of extratropical systems, in which currents can be closely approximated by their geostrophic values. The use of the quasi-geostrophic approximation effectively filters out solutions corresponding to high-speed atmospheric gravity waves. This approximation is not accurate in situations in which *non-geostrophic* currents play an important advective role, *e.g.*, around the frontal discontinuities. Although modern computers allow a fully non-geostrophic description of most geophysical fluids, quasi-geostrophic approximation remains a fundamental tool of theoretical research and is used for practical applications in everyday life: every time we *read* an atmospheric isobaric map in terms of wind flowing along the isobars we are, in fact, using the notion of geostrophic balance.

## Applications of Geophysical Fluid Dynamics: weather and climate

Given the nature and the impact on human society of their research, environmental scientists have had access to powerful computers since the early days of numerical computing. Actually, in the late '40s, the first large-scale application of automatic computing consisted in the first numerical weather forecast, based on greatly simplified equations, which was proposed by Von Neumann and mainly devised by Charney. Since the late '50s, the US technical services have been using computer-assisted numerical integration of relatively accurate equations descriptive of the physics of the atmosphere to routinely produce publicly available weather forecasts.

Geophysical fluids numerical models are usually divided into limited area and global models, the fundamental difference being that the former require lateral boundary conditions (at the boundaries of the integration domain) that have to be continuously provided by either observations or, more typically, larger-scale models, while the latter essentially perform *quasi-autonomous* integrations - obviously, still there is external forcing. In geophysical fluids model, the three-dimensional fields are discretized on a lattice, and the temporal evolution is discretized through the adoption of a time-step. The adoption of a finite spatial and temporal resolution implies that the physical processes occurring on a space and/or time scale smaller than the actual resolution of the model can be taken care of only with approximate parameterizations relating such processes to coarser grain information. The parameterization of the so-called *subgrid* processes is usually heuristic and devised *ad-hoc* by statistical interpretation of observed or model-generated data.

Numerical modeling options strongly rely on the available computer power, so that the continuous improvements in both software and hardware have permitted a large increase in the performances of the models and at the same time an impressive widening of their horizons. On one side, the adoption of finer and finer resolutions has allowed a more detailed description of the large scale features of the dynamics, and, more critically, a more direct physical description of a larger set of processes, thus limiting the need for parameterization procedures, which, where needed, have become more accurate. On the other side, it has been possible to implement and then refine the coupling between models pertaining to different systems having a common boundary, such as the atmosphere and the ocean. The main purpose of such procedure is to study in detail the time-space scales of the interactions between the different systems and reframe such interactions in the more intrinsic terms of internal feedbacks of the compounded system.

Today, computer simulations probably constitute the best *laboratory instruments* for environmental scientists, since on one side permit the testing of theories, at least in the previously exposed *weak* sense, and on the other side provide stimulations for formulating new hypotheses. It is notable that the present most powerful computer of world has been engineered in order to simulate geophysical fluid dynamics with the particular scope of providing the most detailed simulation of the present and past climate.

When considering complex systems like those investigated by the environmental sciences, there is a clear distinction between providing efficient descriptions of the local and of the global properties in the phase space of the system. The local properties of the phase space are addressed by considering dynamic equations able to represent the short-time deterministic evolution of the system. The study of the global properties of the phase space entails considering dynamic equations whose goal is to represent the statistical properties of the system. The efficient description of such properties requires very long integrations of the equations. In order to clarify these well-distinct - albeit intersecting - perspectives, we will specifically discuss the problem of weather forecast and of climate simulation.

**Weather forecast**

The task of weather forecast is to provide the best description of the short-time evolution of the relevant fields descriptive of the state of the atmosphere – wind, temperature, pressure, humidity, precipitation – as represented on a 3D lattice.

Weather forecast is addressed to describing and predicting the instantaneous values of the main atmospheric variables. Weather forecast models use in the equations quite a number of *ad hoc* assumptions and parameterizations that have empirically proved their reliability for that purpose. Since the aim of these models is to be as precise as possible in the local sense, they do not necessarily need to obey global constraints, *e.g.* energy conservation, which are obviously fundamental in terms of Physics, but not necessarily relevant when a small neighborhood of the phase space is considered.

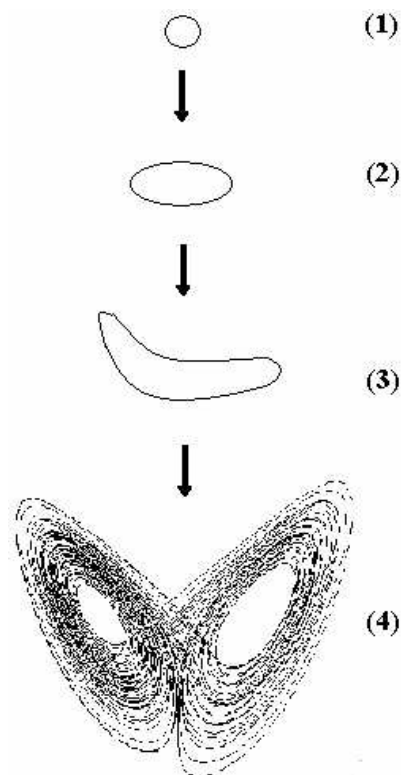

**Figure 5: Scheme of the loss of predictability for ensemble forecasting. The initial Gaussian distribution of the ensemble (1) is distorted (2), and then loses its Gaussian shape (3); eventually the ensemble points populate the attractor of the system.**

The time range of such forecasts extends up to the deterministic predictability horizon of the atmosphere, which can be estimated in about 7-10 days. It is important to emphasize that the actual spread of the model predictions can be estimated by using Monte Carlo techniques on the initial state of the system. Such approach is termed *ensemble forecasting* and relies on sampling the phase space of system by integrating forward in time a set of initial conditions. The initial conditions are suitably selected in order to represent the unavoidable uncertainty on the initial state of the system.

**Climate simulation**

The climate consists of a set of statistical properties - in the time and/or space domains - of quantities that describe the structure and the behavior of the various parts of the climatic system.

The goal of climate models is to represent as well as possible such statistical properties, thus aiming at a precise picture of the main properties of the phase space attractor of the system. In the case of past or present climate change studies, models try to assess how such statistical properties change with time. In the case of climate modeling, the relevant uncertainties are not related to the initial conditions of the system but rather to the very structure of the model. Since *ab-initio* formulations based on first principles are not possible and since we are anyway using with relatively coarse-grain discretization (the best available models have average resolutions of 100 Km in the horizontal and 200 m in the vertical directions), the presence of structural uncertainties, which are equivalent to uncertainties in the attractor's shape, is unavoidable.

A critical point in climate modeling is that, depending on the time scale of interest and on the problem under investigation, the relevant active degrees of freedom, which need the most careful modelization, change dramatically. For relatively short time scales (1-100 years) the atmospheric degrees of freedom are active while the other sub-systems can be considered essentially *frozen*. For longer time scales (100-1000 years) the ocean dominates the dynamics of climate, while for even longer time scales (1000-10000 years) the ice sheet changes are the most relevant factors of variability. Therefore the very structure of climate models is gauged with respect to the purpose of the specific study and, depending on this, the models may be formulated in totally different ways.

Lately, a relevant strategy, which may be termed *statistical ensemble forecasting*, has been proposed for dealing with the delicate issue of structural uncertainties, especially in the context of the study of the future projections of climate change, even if in principle this is not its only application. The main idea is to consider a climate model and apply Monte Carlo techniques not only on the initial conditions, but also on the value of some key uncertain parameters characterizing the climate *machine*. The resulting statistics (and change of the statistics) is expressed as a set of probability distributions rather than as a set of values.

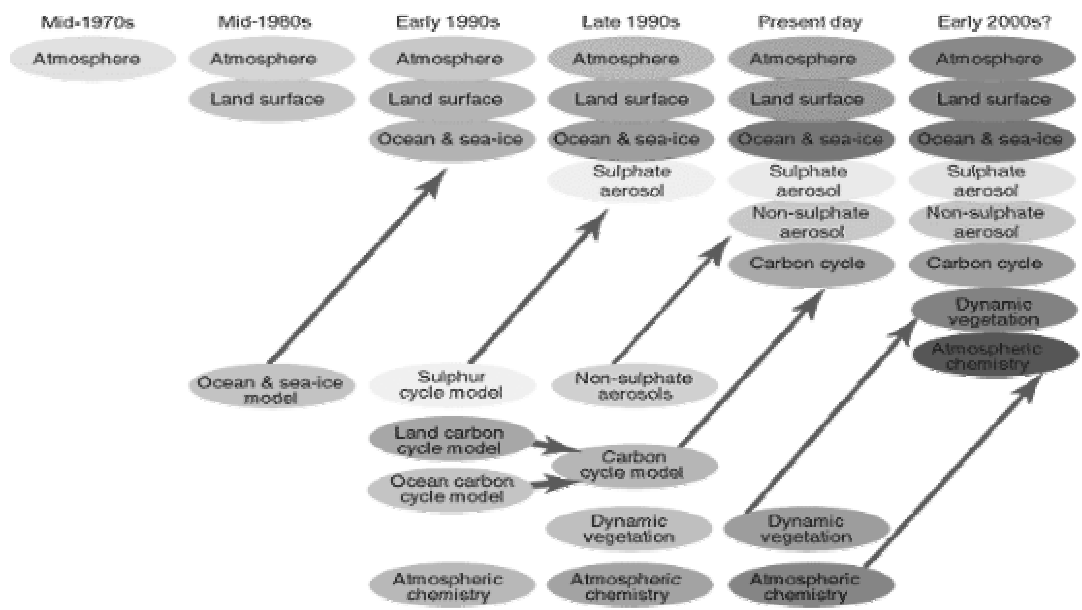

**Figure 6: Evolution of the complexity of climate models in terms of representation of the main processes; the red lines indicate the integration of specific models into climate models. From IPCC 2001**

## Statistical inference

Another prominent approach to the problem of providing an efficient description of the main properties of the environmental systems consists in trying to infer information directly from observations. Such an approach constitutes a natural and necessary complement to the theoretical and model-based analyses, which unavoidably suffer of the above discussed uncertainties. A classical way of gathering information on the investigated system from the actual measurements consists in applying techniques that fall into the wide chapter of *statistical inference*. Environmental Science is rich of such applications, some of which are quite well known to a wide public. In the last two decades, the outstanding example is that of *Global Warming*. In statistical terms, the problem consists in trying to detect signs of anthropogenic changes (expressed in terms of trends) of local as well as global measures of relevant climatic variables. In particular, the average global surface temperature is – somewhat anthropocentrically - usually considered as *the* variable descriptive of such change. The general problem is quite abundantly documented in the reports of the International Panel on Climate Change (IPCC).

What is the concrete appearance of surface temperature time-series? The presence of variability on a vast range of time scales, or, in other terms, of a rather composite spectrum, implies that suitable filtering procedures are needed in order to detect changes with time of the signal. We next present an example of how statistical inference can be applied to characterize the climatology of a system.

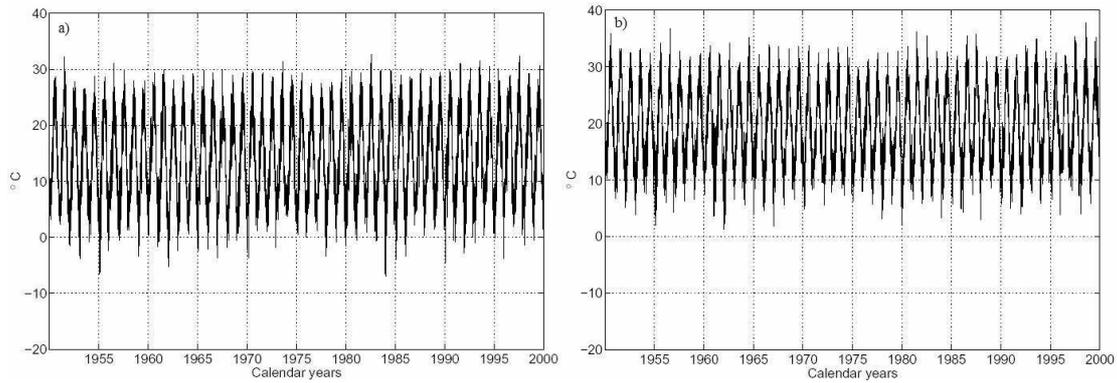

**Figure 7: Daily records (1951-2000) of maximum temperature for northern (a) and southern (b) Italy. The signature of the seasonal cycle is evident.**

In figure 7 we show the 1951-2000 daily maximum temperature records for two synthetic stations representative of northern and southern Italy. These data have been obtained after a suitable homogenization process of few dozens of real meteorological stations in each of the two parts of Italy. Apart from the signature of the seasonal cycle, the signal features an evident noisy structure. When the higher frequency components are removed and the inter-annual components only are retained, we may observe an increasing trend for both records, which is consistent with the large scale picture for the last 50 years. The observed warming trend can be proved to be statistically significant. But, instead of focusing on the long-term changes in the value of the signal, we are often interested in detecting changes in the properties of the seasonal cycle, defined in terms of phase and amplitude of the (1 year)$^{-1}$ component of the spectrum. Such a problem may be approached by providing a time-dependent estimate of the spectrum, obtained with a *shifting window* Fourier analysis, and then performing a statistical analysis of the various estimates of the *locally defined* (in time) seasonal signal.

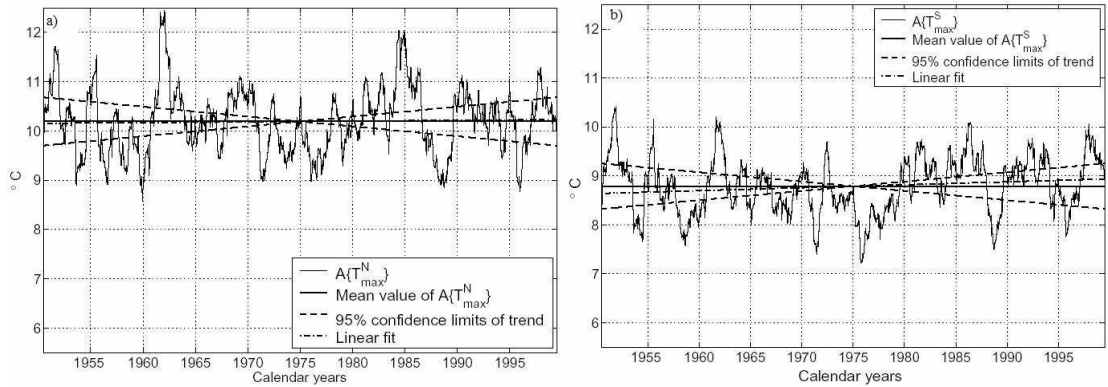

**Figure 8: Amplitude of the seasonal signal** $A\{T_{max}^{N/S}\}$ **for the maximum temperature of northern (a) and southern (b) Italy. No statistically significant trend can be detected in either case.**

In the case of the records shown in figure 7, we have that the amplitude and the phase of the signal, albeit characterized by a relatively large variability, do not have statistically significant trends. In figure 8 we present the results for the amplitude of the signal: the linear trend does not have a statistically well-defined sign for both records. Similar situation occurs for the phase of the signals. From these results we can infer that the seasonal cycle has not undergone relevant long-term changes. Apart from the analysis of the variability of the signals under examination, we can infer from the average properties some basic climatic differences between the two systems, which in this case are defined in terms of geography only. First we have, quite trivially, that northern Italy is colder, since its average temperature is notably smaller (figure 7). This is mostly due to latitudinal effects. Second, northern Italy has a more continental climate, since its seasonal cycle has larger average amplitude (figure 8). As a first approximation, we can interpret this property an effect of the smaller effectiveness of the air-sea thermal exchanges in smoothing out the effect of the annual changes in the intensity of the incoming solar radiation.

**Diagnostics**

Another relevant area of matching between quantitative and qualitative approaches to the study of environmental systems is that of diagnostics of the complex space-time structure of fields based on indicators derived from measure theory. It is well known that, *e.g.*, the definition of the *structure* of precipitation, land cover, *etc.* can be given in terms of estimators of the fractal measure derived from observations. In fact, the very concept of *fractal dimension* has been elaborated also in the effort of *measuring* territorial systems, with the classical example given by Mandelbrot for the evaluation of the length of the southern coast of England.

Let us consider standard *precipitation*, which is defined as the quantity of atmospheric water reaching the Earth's surface in unit time. Such a quantity is the final result of an extremely long and complex chain of events, virtually covering all the space-scales, from molecular water to the global scale of cyclonic areas, and the time-scales from the fractions of second of enucleation to the life-time of weather perturbations (days). Of course, other modulations exist at longer time scales (like the seasonal oscillation), but these are to be considered variation of external forcing rather then part of the precipitation process in itself. Therefore, any series of precipitation observations shows *structure* virtually at all scales. One way of measuring such structure is to make use of the so called *box counting*, which is very instructive since it makes reference to a basic definition of dimension in terms of the statistics of *occupation* of the considered parameter space. When investigating the statistical properties of homogeneous, stationary, and isotropic rainfall, a more appropriate definition of structure functions can be given through the moments of the integral measures of precipitation. Let us consider a positive random field $P(x), P(x) > 0$, defined in the set $x \in [0, L]$. Without loss of generality, *P(x)* is normalized to 1, *i.e.*:

(5)
$$\int_0^L P(y)dy = 1$$

for any realization. Our primary interest is to discuss the scaling properties of the structure function:

(6)
$$\int_x^{x+r} P(y)dy \equiv \mu_x(r)$$

$P(x)$ is said to display anomalous scaling properties if:

(7)
$$\left\langle \left[\mu_x(r)\right]^q \right\rangle \sim r^{\xi(q)},$$

where $\xi(q)$ is a nonlinear function of $q$. The scaling exponents $\xi(q)$ are also referred to as *multi-fractal* exponents of $P(x)$. The random field is assumed to be ergodic, so that in the previous equation $\langle \ \rangle$ stands both for $x$-average ($x$ being either a space or time coordinate) and ensemble average. It follows that we must require also stationarity and homogeneity. Furthermore, if the field is embedded in a space of dimension $d>1$, the field is also assumed to be isotropic, *i.e.* the scaling properties depend only on the modulus of $r$.

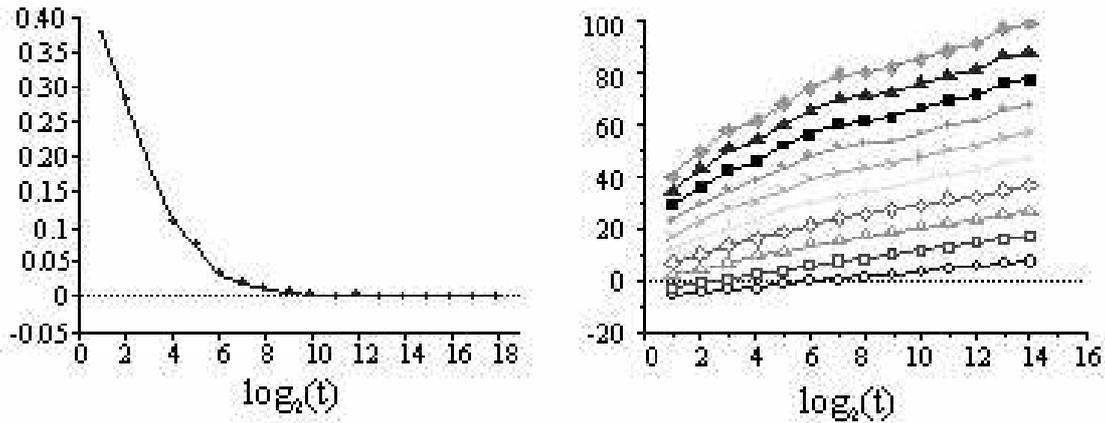

**Figure 9: Time-structure of the precipitation at a rain-gauge in the area of Rome (1 minute interval of sampling from 1992 to 1998). On the left, autocorrelation. On the right, sequence of moments in increasing order from bottom up. The time scale in abscissa is in powers of 2 minutes.**

In figure 9 we present some statistical results for the precipitation in Rome. We learn from correlation (the station of Rome is quite typical in this respect) that precipitation is characterized by a limited (with respect to other atmospheric variables like, for example, the surface temperature considered above) autocorrelation. The scale of decay of correlation is of the order of hours. A simple fractal structure would be represented in the logarithmic diagram of moments by a straight line. Apart from modulations due to sampling, we see that the higher moments show exponent dependence on scale and, therefore, indicate a multi-fractal distribution. The meaning of the *inflection* around $2^8$ m ~ 4 h is that organized dynamical processes (like the formation of *rain bands*) are operating at larger time scales, while the incoherent, micrometeorological processes dominate at smaller scales. This kind of analysis can be performed also in space (although this requires an extensive observational coverage that is rarely available) and produces similar results. Other statistical approaches are possible, in particular for what concerns the parametric distributions that are used in order to represent the above shown type of behavior. As we have seen, such an analysis can identify statistical features revealing aspects of the structure and the dynamics of the systems in question. Another use is for operations of *upscaling* and *downscaling* the observed fields.